\documentclass[conference]{IEEEtran}
\input epsf
\usepackage{graphicx}
\begin{document}

%\title{Submission Format for IMS2014 (Title in 24-point Times font)}
% If the \LARGE is deleted, the title font defaults to  24-point.
% Actually, 
% the \LARGE sets the title at 17 pt, which is close enough to 18-point.
%+++++++++++++++++++++++++++++++++++++++++++
\title{\LARGE Serverless Architecture for Bulk Email Management}
\author{\authorblockN{Bazaru Priyatham Sai Chand}
\IEEEauthorblockA{Dept. of Computer Science\\
Mahatma Gandhi Institute of Technology\\
Hyderabad, Telangana\\
Email: bpriyathamsaichand\_cse180508@mgit.ac.in}
}
\maketitle

\begin{abstract}

 Sending emails in large quantities can be tedious considering free services do not cover bulk email and paid services can be costly and are not easy to customize. Traditional email client used for basic emailing services fail to be useful in larger volumes of emails to target people or spread information to consented individuals. This paper proposes a serverless architecture to tackle such problems by using one such offering from the Amazon Web Services(AWS) which can be easily replaced by a software architects choice of service. The constraints help to make an architecture using components that can fit most of the needs of a serverless backend and extend it to scenarios such mobile notifications, One Time Password (OTP) systems or other means of communication to minimize single point of failure and also decrease the dependency on physical servers for such operations offering a comparable solution within the cloud. The architecture proposed is tested to find the time taken to send the emails of various quantities and see how it affects the cost. The architecture was successful able to send multiple emails in a quick and single invocation and has demonstrated a higher level of scalability compared to conventional methods.\end{abstract}
%\IEEEoverridecommandlockouts
%\begin{keywords}
%Ceramics, coaxial resonators, delay filters, delay-lines, power
%amplifiers.
%\end{keywords}

% for peerreview papers, inserts a page break and creates the second title.
% Will be ignored for other modes.
\IEEEpeerreviewmaketitle

\section{Introduction}
% no \PARstart
Email has become one of the most important means of communications.Sending emails through conventional software like Gmail is limited to a mere thousands which is not sufficient to fit into the business needs and managing the insights of sent emails is crucial to companies to further enhance their approach.\par

Serverless computing is any computing platform that hides server usage from developers and runs code on-demand automatically scaled and billed only for the time the code is running\cite{castro2019rise}.This reduces the time and knowledge required to deploy pieces of software while offering significant reduction in pricing compared to hosting the software on dedicated hardware.Various cloud platforms such as AWS, Microsoft Azure and Google Cloud Platform have adapted to this way of thinking making it easily accessible.A robust serverless architecture for sending bulk emails of the magnitude of thousands to millions using scalable model which is implemented using AWS lambda\cite{awslambda} and Simple Email Service(SES)\cite{awsses} for the SMTP protocol. Challenges such as able to trace the bounce emails as well as the statistics for the emails sent. The prototype is built using the React framework along with underlying services from AWS. These tightly integrated services offer the best way to simulate the architecture and help in sending customized for every recepient based on email templates along with storing bounce emails. \par

The effectiveness of the scalability offered by the architecture is analyzed using X-ray traces to simulate and test the bottlenecks and how to overcome them is discussed in the paper.With serverless technology, the cloud provider abstracts away the server management, provisioning servers with fine granularity, on demand, and with a pay-per-use model\cite{eismann2020serverless}.

This paper proposes an architecture for sending and management of the emails through serverless application using the AWS services. The paper outlines the accuracy and the reliability of the design and future aspects for this architecture.
%\subsection{Subsection Heading Here}

\section{Overview of bulk email management system}
The bulk email management system is used to send emials of magnitude in thousands and more and broadcast the message to multiple email addresses with a single operation. All the required fields such as the senders' email addresses, subject, the html body and to addresses are selected and the sending is initiated with a button. Upon successful sending to the API endpoint the emails are processed and send through the AWS simple email services to all the users and the emails that were bounced are returned after the sending operation.It makes use of a Javascript based framework called ReactJS to input the fields and send it to the API hosted on the AWS service where an array of services are used to scale and perform the sending and retrieving of bounces.

%[22]{L}{0.45\textwidth}
\begin{figure*}[h]
	\centering
\includegraphics[width=.8\textwidth]{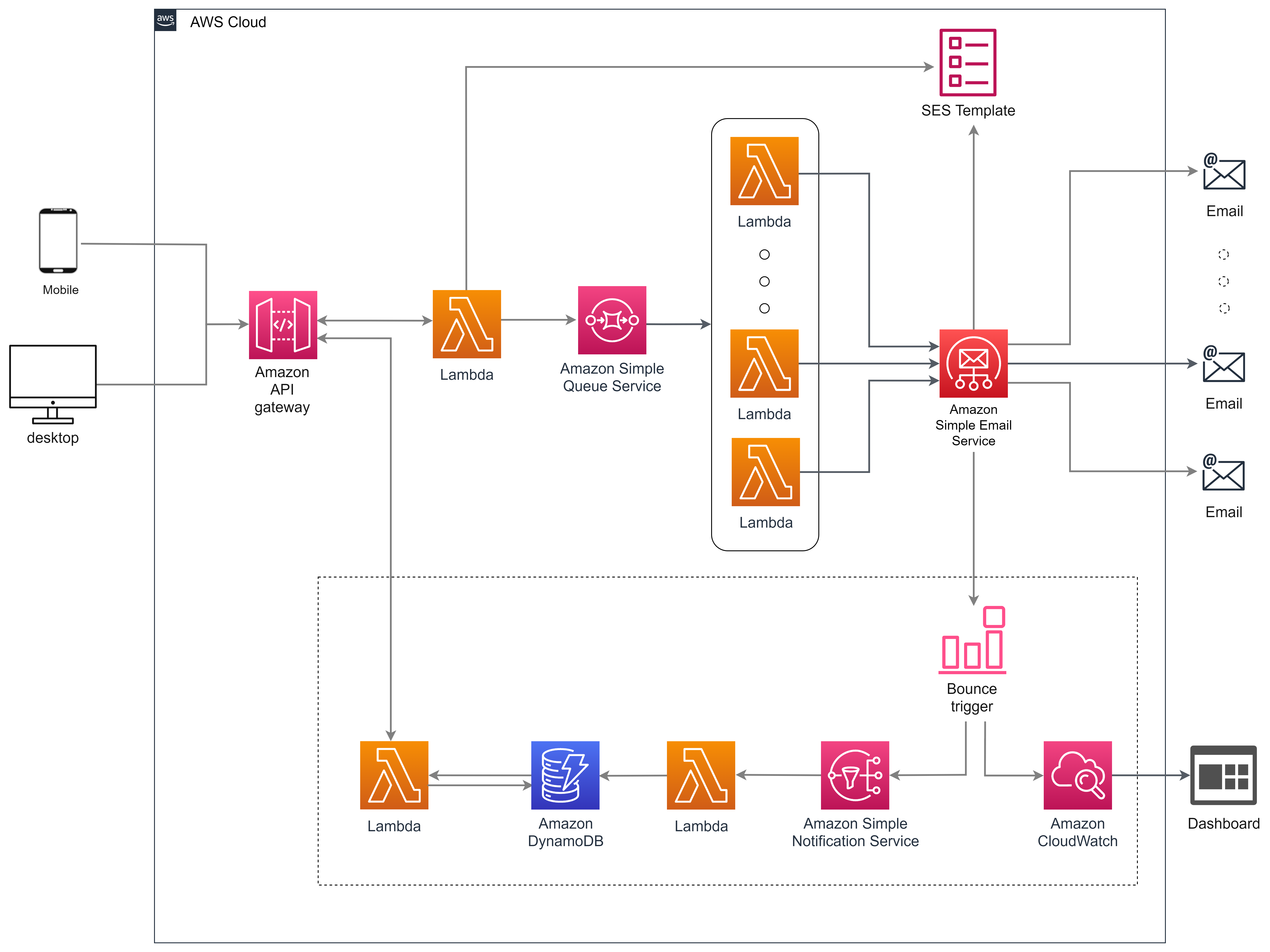}
	\label{fig:fig1}
	\caption{Architecture for bulk email management}
\end{figure*}

\section{Architecture Design}
The architecture demonstrated in the Fig 1 with the main functionalities include sending the emails and tracking the bounces. 
Initially, all the data sent to the lambda using the API gateway which handles the request and response for the whole operation. Upon reaching the lambda, JavaScript Object Notation (JSON) objects from the string parameters preprocessed, the html part of the email along with the subject made into SES templates accessed by other lambdas and the emails batched no more than 50 emails per batch and placed in a Simple queue Service\cite{awssqs} queue. The ordering of the messages is important hence, we use a First In First Out queue, which triggers another lambda that sends emails to all the 50 recipients using a templated email function by accessing the template that we created earlier. We set a flag to mark the end of the batches deleting the template by the end of the execution.

The emails use a configuration set that triggers events such as bounce,complaints or seen to other services for logging and other analysis. The configuration set sends all the bounce receipts to SNS which triggers a lambda function that stores the receipts useful information as a document in a NoSQL database such as dyanamoDB once all the emails sent and all bounce receipts stored in the database. Another lambda triggers using an API retrieves all the records and sends the bounced emails as response to API gateway which forwards it the application shown to the user.

The emails that are bounced are identified by the configuration set and are sent to the SNS with email address and the type of bounce which invokes a lambda trigger that stores the Simple Notification Service(SNS)\cite{awssns}notifications after filtering the message into the dynamo DB.

All the bounced emails are automatically updated to a email supression list which saves the names of bounced emails and accepts them but doesn't send the emails after it is declared bounce once. Thereby, saving the bounce rate of the service.

Additionally, All the other statistics are sent to Cloudwatch metrics which can be used to analysis the statistics of sending emails at different time frames and track the bounce rate and can be useful in finding out the effectiveness of the campaign by the number of people who opened the email.

The architecture keeping in mind the constraints on SES service. The advantages of this is two folds. It helps to work within the limits of the service. Morever, these constraints can be fine tuned to deal with the limits of SMTP server if an independent server or other services are used.

\section{Constraints}

The architecture is based a few constraints from the SES service and other bottlenecks that affect the sending of emails. The SES through which the emails are initially sent through sandbox to test and later needs to be upgraded by the AWS authority to permit the user to request increase in the number of emails to be sent. For the purpose of this architecture 50,000 emails have been requested which was granted by the authority.

\begin{itemize}
	\item 14 emails/second
	\item 50,000 emails per day
	\item request data 5kb
	
\end{itemize}
There are two performance monitoring services
available to AWS application developers: CloudWatch
and
X-ray.
\section{Analysis}
Basic message transfer from Author to Recipients is accomplished by using an asynchronous store-and-forward communication infrastructure in a sequence of independent transmissions through some number of mail transfer agents\cite{rfc5598}. Two of the most common ways to analyze the architecture was to consider the time required to send all the emails and the cost associated with it. Both of the metrics are analysed in the following section. 
\subsection{Performance}
There are two performance monitoring services available to AWS application developers: CloudWatch and X-ray.\cite{lin2018tracking}
 The application is analyzed to see the time required to send the emails and the time taken to run the functions between the request and response to the API endpoint. This can be seen through the cloudwatch\cite{awscloudwatch} logs and X-ray to find out the issues with latency over various regions and finding the best regions and how to optimize the functions.In this case, we used X-Ray as a monitoring and display service that automatically samples the entry and exit of function instances, called segments, using unique trace identifiers \textbf{\texttt{(trace\_id)}}\cite{lin2018tracking}\cite{awsxray}.\par

 Here we take the case of sending multiple emails and the time taken is sampled over 30 data points to represent in Figure 2. The time is calculated by considering the cloudwatch logs that can be used to determine the time taken to execute the last function of lambda as show in Figure 1 to send the email and the time leading to that function from the start of the API gateway request log that is stored in the logs. This experimented several times to cancel out the variations of cold start that is a common phenomenon in serverless functions. All the calculations are done to avoid any overlap between different invocations and only a single request is sent within the same region.

\begin{figure}[h]
\includegraphics[width=.45\textwidth]{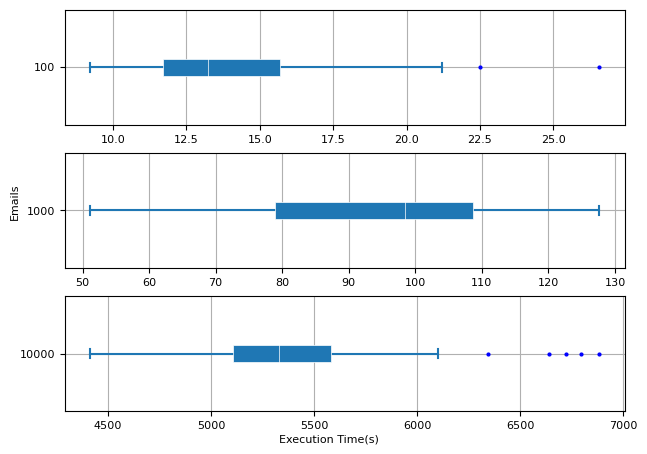}
	\label{fig:}
	\caption{Time taken to execute and return the response to the number of emails that are sent.}
\end{figure}

The number of emails sent is directly proportional to the time taken. The time taken for 10,00 emails averaging at around 30 minutes is sufficient and satisfactory considering that the time taken is to derived from the last email that is sent in the list of emails.  \subsection{Cost}
The cost of the services can also be a motivating factor to implement such an architecture. In our use case The following architecture can be tested with the AWS free tier and later expanded to the needs of the users. With \$0.01 dollars per 1,000 emails but one needs to be wary about the associated services like the SES, SNS and the cloudwatch metrics to analyze the emails add up to cost but since they are calculated based on the GB of data it needs to pre-process the rates will not exceed the cost of emails.

Here we take the example, provided by the SES platform to estimate the cost required to implement the architecture.

You use Amazon SES to send about 250,000 emails per month. You receive 1,000 emails per month. You don't use dedicated IP addresses. Every message you send and receive is 32KB in size which results in a total of \$25.98 per month \cite{sespricing} which is significantly less than competitors such as SendGrid or MailChimp who offer their own SMTP server or schedule emails to fit into the constraints of other providers to carry out email campaigns. \par

Amazon provides sample pricing calculations,but as your workload varies, so will the billing.\cite{eivy2017wary} The SQS requests can exceed the free tier if not monitored carefully and add up to additional costs for the next 1000 requests or more based on the usage of the system.

%\begin{table*}[h]
%	\centering
%	\begin{tabular}{lllll}
%		\toprule
%		\multicolumn{3}{c}{Incoming}                   \\
%		\cmidrule(r){1-3}
%		S no. & Type of charge     & Quantity     &  price  & total amount in dollars \\
%		\midrule
%		1 & Outgoing &250,000 emails & 0.001 & 25     \\
%		2 & Outgoing data & 8GB &  0.12& 0.96 \\
%		\bottomrule
%	\end{tabular}
%	\label{tab:table}
%\end{table*}

\section{Conclusion}
To conclude, the architecture has been effective in sending bulk emails of various volumes with runtime that can be considered as sufficient or optimal and cost-effective. The primary goal of the architecture to be able to manage the entire system without a server being provisioned directly i.e. serverless is achieved and is promising to further develop and enhance its capabilities.  The statistics displayed can be used for reach through the customers and the key idea of the architecture is to make use of serverless functions and services to manage emails with an optional fronted interface which was the case in this paper. Thus, making it a fast application to deliver emails to the users' inboxes in a short period of time. Without provisioning any IP addresses and acquiring the compute power needed to send the emails beforehand thereby making it more effective and possesses a higher level of abstraction that the conventional method.

\bibliographystyle{IEEEtran.bst}
\bibliography{IEEEabrv,./paper_refs}
%\begin{thebibliography}{1}

%H.~Kopka and P.~W. Daly, \emph{A Guide to {\LaTeX}}, 3rd~ed.\hskip 1em plus

%\bibitem{lamport} L. Lamport, \emph{ {\LaTeX} A Document Preparation
%  System}, Reading, Mass: Addison-Wesley, 1994.

%\bibitem{knuth} D. E. Knuth, \emph {The \TeX book}, Reading, Mass.:
%  Addison-Wesley, 1996.

%\end{thebibliography}
\smallskip
\end{document}